\documentclass[aps,pra,reprint,nofootinbib,superscriptaddress]{revtex4-2}

\usepackage[T1]{fontenc}
\usepackage[utf8]{inputenc}
\usepackage{lmodern}
\usepackage{amsmath,amssymb,mathtools,bm}
\usepackage{graphicx}
\usepackage{booktabs}
\usepackage{array}
\usepackage{xcolor}
\usepackage{makecell}

\usepackage[colorlinks=true,linkcolor=blue,citecolor=blue,urlcolor=blue]{hyperref}

\graphicspath{{figures/}}

\newcommand{\Gbb}{\mathcal G_{BB}}
\newcommand{\Gsczz}{\mathcal G^{\mathrm{sc}}_{BB,zz}}
\newcommand{\rte}{r_{\mathrm{TE}}}
\newcommand{\rslab}{r_{\mathrm{TE}}^{\mathrm{slab}}}
\newcommand{\kb}{k_\perp}
\newcommand{\dd}{\mathrm d}
\newcommand{\nth}{n_{\mathrm{th}}}
\newcommand{\RR}{\mathbf R}

\begin{document}

\title{Finite-frequency magnetic common baths in ferromagnetic planar cavities}

\author{O. J. Kwon}
\affiliation{Independent Institute for Theoretical Studies, Yongin, Republic of Korea}
\email{ojkwon@iits.re.kr}


\begin{abstract}
We formulate the finite-frequency magnetic common-bath problem for two spin probes in a ferromagnetic planar cavity. The probes couple to the retarded magnetic Green tensor of the cavity, whose imaginary and real parts determine the collective decay rate \(\gamma_{12}\) and the coherent exchange coupling \(\Omega_{12}\), respectively. We evaluate these quantities for a finite-thickness scalar transverse-electric (TE) cavity channel formed by ferromagnetic films on nonmagnetic conducting substrates, using a Polder-type scalar permeability \(\mu_\perp(\omega,B)\) as the magnetic input. The normalization is fixed by the free-space magnetic-dipole decay rate. In the \(\omega\to0\) limit, the Green-tensor kernel reduces to the magnetostatic response, and the finite slab retains its static TE reflection amplitude. For a \(t=200\,\mathrm{nm}\) film with a representative Ni-like parameter set, the GHz transition of a spin probe located at the midplane of a micron-scale cavity samples the spectral density of the body-assisted magnetic reservoir at millikelvin temperature. At the resonant bias, the cavity-induced contribution \(2|\delta\gamma_{12}^{(zz)}|\) to the collective linewidth splitting in the \(zz\) transition-moment projection is about two thirds of the local single-spin linewidth correction \(\delta\gamma_{11}^{(zz)}\) at \(\rho=3\,\mu\mathrm m\). Together, the linewidth splitting and collective Lamb shift probe the dissipative and dispersive finite-frequency response, respectively, while the shared Green-tensor normalization connects these observables to the static TE coupling-frequency shift.
\end{abstract}

\maketitle

\begin{figure*}[t]
\centering
\includegraphics[width=\textwidth]{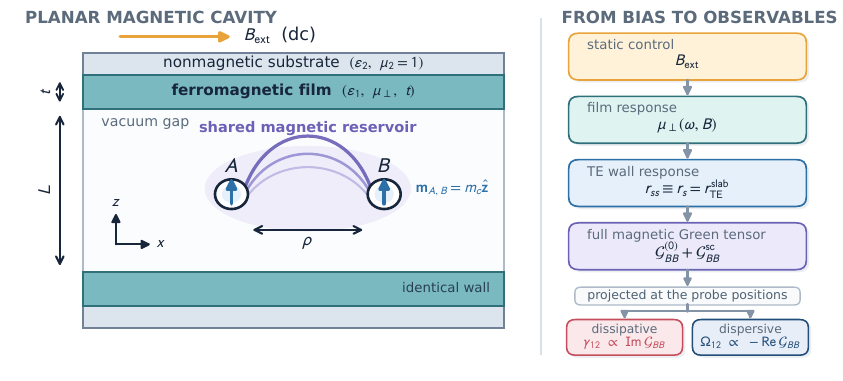}
\caption{Geometry and finite-frequency common-bath mechanism. The left panel shows two probes in the vacuum gap between identical ferromagnetic-film--substrate walls, separated by an in-plane distance \(\rho\) and coupled through a shared magnetic reservoir. The \(z\) axis is normal to the films, the \(x\) axis is chosen along the in-plane probe separation, and the benchmark transition moments are \(\mathbf m_A=\mathbf m_B=m_c\hat{\mathbf z}\). The right panel summarizes the hierarchy from the static bias \(B\equiv B_{\mathrm{ext}}=\mu_0H_{\mathrm{ext}}\) and film response \(\mu_\perp(\omega,B)\) to the TE reflection amplitude and the full retarded magnetic Green tensor. Projection at the probe positions gives the dissipative kernel \(\gamma_{12}\propto\Im\Gbb\) and the dispersive kernel \(\Omega_{12}\propto-\Re\Gbb\). The benchmark figures use \(t=200\,\mathrm{nm}\), nonmagnetic conducting substrates, and the \(zz\) transition-moment projection.}
\label{fig:schematic}
\end{figure*}

\section{Introduction}
\label{sec:intro}

Reference~\cite{Kwon2026Magnetostatic} showed that changing the magnetic state of ferromagnetic films controls the zero-frequency TE boundary response of a planar cavity and produces a static Ising coupling-frequency shift \(\Delta f\) between two localized Zeeman probes in the gap. That result motivates extending the same Green-tensor framework to real frequencies, where the cavity mediates dissipative and coherent coupling between spin probes. Here we carry out this extension and formulate the finite-frequency magnetic common-bath problem for two spin probes in a ferromagnetic planar cavity.

We consider two magnetic two-level probes fixed in the vacuum gap between two ferromagnetic films (Fig.~\ref{fig:schematic}). By reshaping the magnetic-field fluctuations in the gap, the cavity boundaries provide a controlled reservoir shared by the two probes. At GHz frequencies and millikelvin temperature, the probes sample this reservoir at their transition frequency. The off-diagonal projection of the retarded magnetic Green tensor describes the common-bath response. Its imaginary and real parts determine the collective decay rate \(\gamma_{12}\) and the coherent exchange coupling \(\Omega_{12}\), respectively. The bias tunes the ferromagnetic resonance (FMR) through the fixed probe transition frequency. We label the resonant bias setting ON and the detuned settings OFF. We decompose the Green tensor into a free-space part and a cavity-induced scattering correction; the benchmark figures show the latter, whose ON/OFF difference removes bias-independent offsets.

Spin-flip transitions induced by magnetic-field noise near conducting surfaces have been studied in the atom-chip setting~\cite{Jones2003SpinCoupling,Rekdal2004,Scheel2005}; here we derive the two-probe common-bath response for a ferromagnetic planar cavity. In magnetic materials, quantum-impurity relaxometry relates a single probe's relaxation rate to the imaginary part of the dynamical magnetic response~\cite{Flebus2018}. The collective linewidth splitting set by \(\gamma_{12}\) probes the projected magnetic-field cross-spectral density between the two probe positions at \(\omega_{\mathrm{pr}}\), while \(\Omega_{12}\) gives the associated dispersive response.

Our formulation combines macroscopic quantum electrodynamics (QED) in absorbing, dispersive geometries~\cite{Agarwal1975Spontaneous,ScheelBuhmann2008,Buhmann2012,Rapp2025Purcell} with the Dicke--Lehmberg description of collective emission~\cite{Dicke1954,Lehmberg1970a,Lehmberg1970b,GrossHaroche1982,FicekTanas2002}. Related extensions treat emitters coupled through one-dimensional guided channels~\cite{Solano2023Dissimilar,Lee2023Collective}. In a structured environment, collective decay and coherent exchange follow from the imaginary and real parts of the retarded Green tensor, respectively~\cite{DungKnollWelsch2002}. For the present planar magnetic geometry, we construct that tensor from Fresnel scattering amplitudes~\cite{WylieSipe1984,WylieSipe1985,Tomas1995Green} and project the curl-defined magnetic Green tensor onto the probe transition moments~\cite{ScheelBuhmann2008,Buhmann2012}; related Green-tensor treatments of surface-modified collective emission and resonant dipole--dipole interactions appear in Refs.~\cite{Jones2018,Skljarow2022,SvendsenOlmos2023,Araujo2024Cooperative}.

We use a common magnetic Green-tensor normalization for the static coupling-frequency shift \(\Delta f\) and the real-frequency open-system kernels \(\gamma_{12}\) and \(\Omega_{12}\). We evaluate the off-diagonal decay rates and exchange couplings for two probes in the mid-gap region of a finite-thickness ferromagnetic planar cavity, with the FMR bias serving as a spectroscopic tuning parameter. The resulting linewidth splitting and collective Lamb shift probe the dissipative and dispersive two-point magnetic response, respectively.

The paper is organized as follows. Section~\ref{sec:model} defines the hierarchy from material response to observables, the Green-tensor convention, the finite-thickness scalar TE channel, and the magnetic response model. Section~\ref{sec:opensystem} formulates the vector-projected open-system kernels and identifies the \(zz\) projection as a benchmark choice. Section~\ref{sec:numerics} describes the numerical implementation and presents the benchmark figures. Section~\ref{sec:discussion} discusses differential observables, a proposed measurement protocol, and the controlled scope of the scalar calculation.

\section{Model}
\label{sec:model}

\subsection{Hierarchy, geometry, and Green-tensor convention}
\label{subsec:geometry}

Connecting the film response to the two-probe dynamics requires several successive levels of description. We consider the reciprocal TE channel of a ferromagnetic film of thickness \(t\) on a nonmagnetic conducting substrate and describe the film by an effective scalar magnetic response. The static bias enters through the resonance frequency \(\omega_0(B)\). The model retains finite film thickness, Ohmic loss, the FMR pole, and cavity multiple scattering, while omitting TE--TM conversion and the gyrotropic off-diagonal permeability components. The resulting hierarchy is
\begin{equation}
\begin{aligned}
\{\varepsilon_1(\omega),\mu_\perp(\omega,B),t,\varepsilon_2(\omega)\}
&\to r_s(\omega,k_\perp,B)
\to \mathcal G_{BB}^{\mathrm{sc}}(\omega,B),
\\
\mathcal G_{BB}^{(0)}+\mathcal G_{BB}^{\mathrm{sc}}
&\to \{\gamma_{ij},\Omega_{ij}\}.
\end{aligned}
\label{eq:scalar_hierarchy}
\end{equation}
Here \(t\) is the film thickness, \(\varepsilon_1\) and \(\varepsilon_2\) are the film and substrate permittivities shown in Fig.~\ref{fig:schematic}, and \(\mu_\perp(\omega,B)\) is the effective scalar permeability of the film. These material inputs determine the one-wall TE reflection amplitude \(r_s(\omega,k_\perp,B)\), defined in Sec.~\ref{subsec:kernel}. At a fixed probe frequency, the bias dependence of this scalar benchmark enters only through \(\mu_\perp(\omega,B)\).

The two probes occupy a vacuum gap of width \(L\) between two identical ferromagnetic films, as sketched in Fig.~\ref{fig:schematic}. Their positions are \(\mathbf r_A=(\bm\rho_A,z_A)\) and \(\mathbf r_B=(\bm\rho_B,z_B)\), with in-plane separation \(\rho=|\bm\rho_A-\bm\rho_B|\).
The finite-frequency problem is formulated in terms of the magnetic Green tensor used in macroscopic QED~\cite{ScheelBuhmann2008,Buhmann2012},
\begin{equation}
\Gbb(\mathbf r,\mathbf r';\omega)
=
\nabla\times\mathbf G(\mathbf r,\mathbf r';\omega)
\times\overleftarrow{\nabla}',
\label{eq:GBBdef}
\end{equation}
where \(\mathbf G\) is the retarded electric-field dyadic Green tensor~\cite{ScheelBuhmann2008,Buhmann2012}. The curls at the observation and source points convert the electric dyadic Green tensor into the magnetic-field response relevant to magnetic-dipole transitions. We define \(\Gbb\) without the prefactor \(\mu_0\); the factors of \(\mu_0\) required for physical rates and frequency shifts are written explicitly in the formulas below. In homogeneous free space, the isotropic imaginary part of this tensor at coincident points is given by
\begin{equation}
\Im \mathcal G^{(0)}_{BB,ij}(\mathbf r,\mathbf r;\omega)
=
\frac{k_0^3}{6\pi}\delta_{ij},
\qquad k_0=\frac{\omega}{c},
\label{eq:free_norm_intro}
\end{equation}
as shown in Appendix~\ref{app:normalization}.
For a magnetic transition moment \(\mathbf m=m\hat{\mathbf e}\), substituting Eq.~\eqref{eq:free_norm_intro} into the magnetic-dipole decay rate \(\Gamma_m=(2\mu_0/\hbar)\,\mathbf m\cdot\Im\Gbb^{(0)}(\mathbf r,\mathbf r;\omega)\cdot\mathbf m\)~\cite{ScheelBuhmann2008,Buhmann2012} gives
\begin{equation}
\Gamma_m^{(0)}=
\frac{2\mu_0m^2}{\hbar}\frac{k_0^3}{6\pi}
=
\frac{\mu_0m^2k_0^3}{3\pi\hbar}.
\label{eq:free_decay_intro}
\end{equation}

This fixes the prefactors used below; the separation of the singular local free-space self term from the finite cavity-induced scattering correction is detailed in Appendix~\ref{app:normalization}.

\subsection{Finite-thickness scalar TE channel}
\label{subsec:kernel}

The next step in the hierarchy is to construct the two-interface cavity Green tensor from the one-wall response \(r_s\). Here TE denotes the \(s\)-polarized component of the plane-wave angular-spectrum decomposition of the planar Green tensor~\cite{WylieSipe1984,WylieSipe1985}. The probe orientation enters separately through the projection of \(\Gbb\) onto the transition moments. For each in-plane wave number \(k_\perp\), \(r_s(\omega,k_\perp,B)\) is the corresponding TE reflection amplitude of one vacuum--film--substrate wall as viewed from the vacuum gap. To obtain the reciprocal scalar TE channel, we retain the TE part of the planar scattering Green tensor when evaluating Eq.~\eqref{eq:GBBdef} and use translational invariance in the interface plane. Because the magnetic polarization of a TM plane wave lies in the interface plane, the TM channel has no \(zz\) projection. In a polarization-diagonal stack, this component is therefore carried by the TE amplitude alone. The gyrotropic case is not reducible in the same way, since polarization conversion re-enters through the cavity round trips. In this scalar two-interface channel, repeated round trips form the geometric factor \((1-r_s^2e^{-2\kappa_0L})^{-1}\); the azimuthal integration over \(\mathbf k_\perp\) yields \(J_0(\kb\rho)\), and the \(zz\) magnetic projection contributes the prefactor \(\kb^2/(2\kappa_0)\)~\cite{WylieSipe1984,Buhmann2012}. The scattering projection is
\begin{widetext}
\begin{equation}
\begin{aligned}
\Gsczz(\rho,z_A,z_B;\omega,B)
=&\frac{1}{2\pi}
\int_0^\infty \dd \kb\,\kb\,
\frac{\kb^2}{2\kappa_0}
J_0(\kb\rho)
\frac{\mathcal N(\kb,z_A,z_B;\omega,B)}
{1-r_s^2 e^{-2\kappa_0L}},
\\[1mm]
\mathcal N
=&r_s\left[e^{-\kappa_0\Sigma_z}+e^{-\kappa_0(2L-\Sigma_z)}\right]
+r_s^2\left[e^{-\kappa_0(2L-\Delta z)}+e^{-\kappa_0(2L+\Delta z)}\right],
\end{aligned}
\label{eq:planarGBB_scalar}
\end{equation}
\end{widetext}
where \(\Sigma_z=z_A+z_B\) and \(\Delta z=|z_A-z_B|\). The four exponentials correspond to source-image separations \(\Sigma_z\) and \(2L-\Sigma_z\) for the single-reflection paths and \(2L-\Delta z\) and \(2L+\Delta z\) for the double-reflection paths. Expanding the denominator in powers of \(r_s^2e^{-2\kappa_0L}\) shows that the terms proportional to \(r_s\) and \(r_s^2\) generate the odd- and even-reflection branches, respectively.

The retarded branch is defined by replacing \(\omega\to\omega+i0^+\),
\begin{equation}
\kappa_0=\sqrt{\kb^2-(\omega+i0^+)^2/c^2}.
\label{eq:kappa_branch}
\end{equation}
Thus, \(\Re\kappa_0>0\) in the evanescent sector, while \(\kappa_0=-i\sqrt{\omega^2/c^2-\kb^2}\) for \(\kb<\omega/c\). For GHz frequencies and micron-scale separations, the integral is dominated by the near field, \(\kb\sim1/L\gg k_0\), so the propagating sector contributes negligibly to the displayed observables. The branch prescription enforces causality of the retarded Green function.

The finite-frequency kernel in Eq.~\eqref{eq:planarGBB_scalar} has the same scalar multiple-reflection structure as the static cavity. In the present convention, \(\mathcal G_{BB}\) is defined without the vacuum-permeability factor \(\mu_0\), whereas the magnetostatic Green tensor is normalized as the SI magnetic-induction response \(\mathbf B\) to a magnetic-dipole source. The two conventions are therefore related by
\begin{equation}
G^{\mathrm{stat,sc}}_{BB,zz}(0,B)=
\mu_0\lim_{\omega\to0}\mathcal G^{\mathrm{sc}}_{BB,zz}(\omega,B),
\label{eq:static_convention_relation}
\end{equation}
where the same scalar static reflection kernel is used on both sides. Appendix~\ref{app:staticlimit} gives the explicit image-series reduction for the constant-\(r\) static reference. For the finite-thickness slab benchmark, the \(\omega\to0\) limit retains the \(k_\perp\)-dependent slab amplitude \(\rslab(0,k_\perp,B;t)\) in the \(k\)-space integral.

\subsection{Finite film and scalar magnetic response}
\label{subsec:ferro}

\begin{figure*}[t]
\centering
\includegraphics[width=\textwidth]{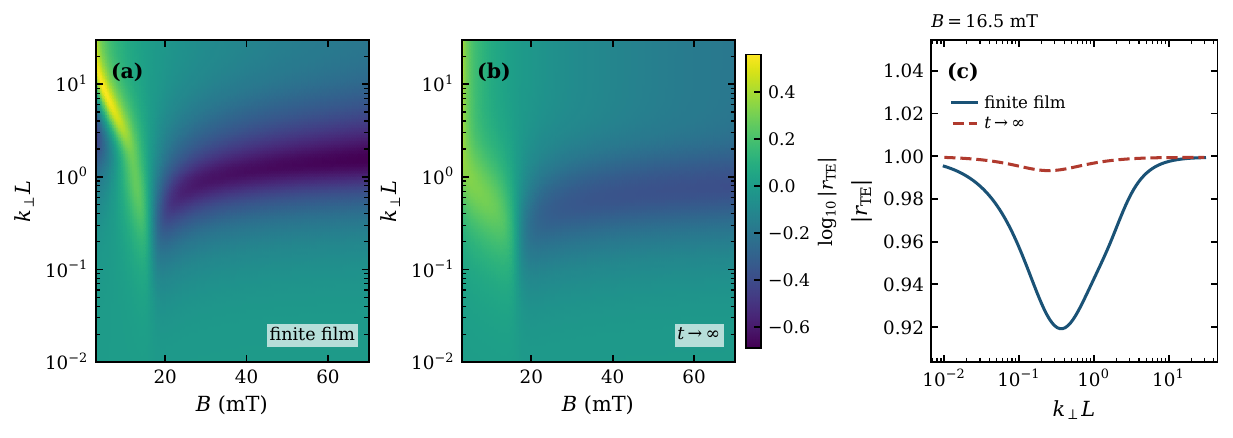}
\caption{Finite-thickness scalar TE benchmark.
(a) Finite-film TE slab amplitude for \(t=200\,\mathrm{nm}\), plotted as
\(\log_{10}|\rte|\) versus \(B\) and \(k_\perp L\).
(b) The corresponding \(t\to\infty\) half-space reference.
(c) A cut at \(B=16.5\,\mathrm{mT}\), where finite thickness reduces
\(|\rte|\) relative to the half-space reference mainly at intermediate
\(k_\perp L\). At large \(k_\perp L\), the finite-film and half-space
curves approach each other.
The quantities shown are calculated from the reciprocal diagonal TE
channel used in the numerical Green-tensor integral.}
\label{fig:reflection}
\end{figure*}

Having expressed the cavity kernel in terms of \(r_s\), we construct the one-wall response as the TE reflection amplitude of a finite-thickness vacuum--film--substrate stack. Region \(0\) is the vacuum gap, region \(1\) is the ferromagnetic film with permittivity \(\varepsilon_1\), and region \(2\) is the substrate with permittivity \(\varepsilon_2\). Only the film is magnetic, with \(\mu_1=\mu_\perp(\omega,B)\) and unit relative permeability in the gap and the substrate. For a local scalar TE channel,
\begin{equation}
\kappa_j=\left[\kb^2-\varepsilon_j(\omega)\mu_j(\omega)\frac{(\omega+i0^+)^2}{c^2}\right]^{1/2},
\qquad \Re\kappa_j\ge0,
\label{eq:kappa_layers}
\end{equation}
and the TE amplitude at an \(a|b\) interface is the standard TE Fresnel reflection coefficient for a planar magnetodielectric interface~\cite{Buhmann2012},
\begin{equation}
r_{ab}^{\mathrm{TE}}
=
\frac{\mu_b\kappa_a-\mu_a\kappa_b}
{\mu_b\kappa_a+\mu_a\kappa_b}.
\label{eq:rab_te}
\end{equation}
In this scalar channel the film acts as a Fabry--Pérot layer for each in-plane wave number. Summing the direct front-surface reflection and the internally reflected round trips gives
\begin{equation}
\rslab(\omega,\kb,B;t)
=
\frac{r_{01}^{\mathrm{TE}}+r_{12}^{\mathrm{TE}}e^{-2\kappa_1 t}}
{1+r_{01}^{\mathrm{TE}}r_{12}^{\mathrm{TE}}e^{-2\kappa_1 t}}.
\label{eq:rslab}
\end{equation}
In Eq.~\eqref{eq:planarGBB_scalar}, we set \(r_s=\rslab\). The half-space reference follows from the limit \(t\to\infty\), where \(r_s\to r_{01}^{\mathrm{TE}}\). When \(\Re\kappa_1t\gg1\), the factor \(e^{-2\kappa_1t}\) in Eq.~\eqref{eq:rslab} suppresses the rear-interface contribution, so the finite-film amplitude approaches the half-space result at large \(k_\perp\).

For the film, we use the low-frequency Drude form with constant conductivity, \(\varepsilon_1(\omega)=1+i\sigma/(\varepsilon_{\mathrm{vac}}\omega)\)~\cite{Buhmann2012}, where \(\varepsilon_{\mathrm{vac}}\) is the vacuum permittivity; the indexed quantities \(\varepsilon_j\) and \(\mu_j\) denote the relative permittivity and permeability of region \(j\). The conducting substrate uses the same functional form, \(\varepsilon_2(\omega)=1+i\sigma/(\varepsilon_{\mathrm{vac}}\omega)\), with \(\mu_2=1\). The film dynamics entering the scalar benchmark are represented by an effective Polder-type single-resonance permeability whose pole is set by the uniform FMR frequency of the linearized Landau--Lifshitz--Gilbert description~\cite{Gilbert2004,StancilPrabhakar2009,Polder1949},
\begin{equation}
\mu_\perp(\omega,B)=
1+
\frac{\omega_M[\omega_0(B)-i\alpha\omega]}
{[\omega_0(B)-i\alpha\omega]^2-\omega^2},
\label{eq:polder}
\end{equation}
where \(\omega_M=\gamma_{\mathrm F}\mu_0M_s\) and \(\gamma_{\mathrm F}/2\pi=28.0\,\mathrm{GHz/T}\).
The replacement \(\omega_0(B)\to\omega_0(B)-i\alpha\omega\) implements phenomenological Gilbert damping with the dimensionless parameter \(\alpha\)~\cite{Gilbert2004,StancilPrabhakar2009}. For an in-plane magnetized film, we use the Kittel resonance condition in the effective-magnetization form~\cite{Kittel1948}
\begin{equation}
\omega_0^{\parallel}(B)=
\gamma_{\mathrm F}\sqrt{B(B+\mu_0M_{\mathrm{eff}})},
\label{eq:kittel_inplane}
\end{equation}
where \(B\) denotes \(\mu_0H_{\mathrm{ext}}\) in tesla. The benchmark permeability follows by setting \(\omega_0=\omega_0^{\parallel}\) in Eq.~\eqref{eq:polder}. The full in-plane gyrotropic susceptibility retains unequal transverse stiffnesses and a nonzero off-diagonal Polder component; incorporating this tensor response requires the full TE/TM reflection matrix of Eq.~\eqref{eq:reflection_matrix}. In the benchmark figures, we set \(\mu_0M_{\mathrm{eff}}=\mu_0M_s=0.617\,\mathrm T\), with \(\mu_0M_s\) taken from the \(293\,\mathrm K\) bulk-Ni saturation magnetization~\cite{Crangle1971}. This illustrative in-plane Ni-like reference neglects any separate perpendicular-anisotropy contribution to \(M_{\mathrm{eff}}\), which would shift the resonance field without altering the scalar Green-tensor construction. Other bias geometries require the corresponding demagnetizing-field form of the resonance condition.

Figure~\ref{fig:reflection} shows the finite-slab response and the \(t\to\infty\) reference. For the benchmark parameters, the in-plane Kittel condition \(\omega_0^{\parallel}(B)=\omega_{\mathrm{pr}}\) is met at \(B=16.58\,\mathrm{mT}\), where \(\omega_{\mathrm{pr}}\) is the fixed probe transition frequency used to sample the Green tensor (Table~\ref{tab:params}). The figures use the rounded resonant value \(B=16.5\,\mathrm{mT}\) for ON and the detuned values \(B=8\) and \(40\,\mathrm{mT}\) for OFF. Within this reciprocal scalar TE benchmark, the field dependence of \(\mu_\perp(\omega,B)\) across this resonance window sets the size and sign of the ON/OFF contrast of the common-bath matrix elements. The Green tensor is evaluated at fixed \(\omega_{\mathrm{pr}}\) throughout; Section~\ref{sec:discussion} treats the case in which the bias also shifts the probe transition.

The scalar Green tensor in Eq.~\eqref{eq:planarGBB_scalar} uses the same reciprocal TE slab amplitude at both cavity interfaces and corresponds to the \(r_{ss}\equiv r_s\) channel of the full planar reflection matrix
\begin{equation}
\RR(\omega,\kb;B)=
\begin{pmatrix}
r_{ss} & r_{sp}\\
r_{ps} & r_{pp}
\end{pmatrix},
\label{eq:reflection_matrix}
\end{equation}
in the TE/TM basis~\cite{Berreman1972}. A full gyrotropic treatment retains \(r_{pp}\) and the TE/TM conversion amplitudes \(r_{sp}\) and \(r_{ps}\), together with the adjoint-tensor definitions of the generalized real and imaginary parts appropriate to nonreciprocal media~\cite{Fuchs2017Nonreciprocal,Franca2025Spectroscopic}.

\section{Open-system kernels}
\label{sec:opensystem}

\subsection{Vector-projected rates and the \(zz\) benchmark projection}
\label{subsec:spectra}

Having specified the film response and the resulting cavity Green tensor, we project this tensor onto the transition moments of two magnetic two-level probes to obtain the dissipative and coherent open-system kernels. The probe operators are
\begin{equation}
\sigma_i^-=|g_i\rangle\langle e_i|,\qquad
\sigma_i^+=(\sigma_i^-)^\dagger,
\label{eq:sigma_def}
\end{equation}
and their free Hamiltonian is
\begin{equation}
H_s=\hbar\omega_{\mathrm{pr}}\sum_{i=A,B}\sigma_i^+\sigma_i^-.
\label{eq:Hs}
\end{equation}
They couple to the medium-assisted magnetic field through the dipole interaction
\begin{equation}
H_{\mathrm{int}}
=-\sum_{i=A,B}\hat{\mathbf m}_i\cdot
\hat{\mathbf B}(\mathbf r_i),
\label{eq:Hint}
\end{equation}
where \(\hat{\mathbf m}_i\) is the magnetic dipole moment operator of probe \(i\) and \(\hat{\mathbf B}(\mathbf r_i)\) is the medium-assisted magnetic field operator at its position.

For the downward magnetic transition moments \(\mathbf m_i=\langle g_i|\hat{\mathbf m}_i|e_i\rangle\), the Green tensor enters the open-system dynamics through the vector-projected scalar kernel
\begin{equation}
\mathcal K_{ij}(\omega,B)
=
\mathbf m_i^*\cdot
\Gbb(\mathbf r_i,\mathbf r_j;\omega,B)
\cdot\mathbf m_j.
\label{eq:Kij_def}
\end{equation}
Within the Born--Markov approximation, the rate matrix is set by the projected magnetic spectral density
\(\mathbf m_i^*\cdot\Im\Gbb(\mathbf r_i,\mathbf r_j;\omega_{\mathrm{pr}},B)\cdot\mathbf m_j\). Its off-diagonal elements (\(i\neq j\)) give the cross-damping terms that, together with the diagonal rates, determine the collective-mode decay rates~\cite{ScheelBuhmann2008,Buhmann2012,DungKnollWelsch2002}. With thermal occupation \(\nth(\omega,T)=(e^{\hbar\omega/k_BT}-1)^{-1}\), emission into the body-assisted field contributes the factor \(\nth+1\), including spontaneous and thermally stimulated emission, whereas absorption from the thermally occupied field contributes the factor \(\nth\). The downward and upward rate matrices are
\begin{equation}
\gamma^{\downarrow}_{ij}=
\frac{2\mu_0}{\hbar}
[\nth(\omega_{\mathrm{pr}},T)+1]
\mathbf m_i^*\cdot
\Im \Gbb(\mathbf r_i,\mathbf r_j;\omega_{\mathrm{pr}},B)
\cdot\mathbf m_j,
\label{eq:gamma_down}
\end{equation}
\begin{equation}
\gamma^{\uparrow}_{ij}=
\frac{2\mu_0}{\hbar}
\nth(\omega_{\mathrm{pr}},T)
\mathbf m_i\cdot
\Im \Gbb(\mathbf r_i,\mathbf r_j;\omega_{\mathrm{pr}},B)
\cdot\mathbf m_j^*.
\label{eq:gamma_up}
\end{equation}
Emission and absorption use Hermitian-conjugate transition moments, so the two contractions are taken in opposite order and coincide in the reciprocal scalar channel used here. A gyrotropic treatment must retain the distinct ordering together with the generalized imaginary part appropriate to a nonreciprocal response.

At \(T=20\,\mathrm{mK}\) and \(\omega_{\mathrm{pr}}/2\pi=2.87\,\mathrm{GHz}\),
\begin{equation}
\frac{\hbar\omega_{\mathrm{pr}}}{k_BT}=6.89,
\qquad
\nth=1.02\times10^{-3},
\label{eq:nth_value}
\end{equation}
so the upward rate is negligible in the benchmark.

The main figures use the component projection \(\mathbf m_i=m_c\hat{\mathbf z}\), so that \(\mathcal K_{ij}=m_c^2\mathcal G_{BB,zz}\). We choose this \(zz\) component because the same projection determines the zero-frequency TE kernel and the Ising coupling-frequency shift \(\Delta f\) in the static calculation~\cite{Kwon2026Magnetostatic}; the two Green-tensor conventions are related by Eq.~\eqref{eq:static_convention_relation}. For a nitrogen-vacancy (NV) center transition, \(m_c=g_e\mu_B/\sqrt2\) is the Cartesian spin-flip matrix element. The physical NV transition moment is transverse to its quantization axis~\cite{Doherty2013}; its circularly polarized contraction and the corresponding component-normalized projection are derived and evaluated in Appendix~\ref{app:transverseprojection}.

\subsection{Master equation, collective decay, and coherent exchange}
\label{subsec:master}

For two identical magnetic two-level systems coupled to a structured macroscopic-QED reservoir, the Born--Markov master equation for a single transition has the collective-emission form~\cite{Lehmberg1970a,Lehmberg1970b,FicekTanas2002,DungKnollWelsch2002}
\begin{equation}
\begin{aligned}
\dot\varrho=&-\frac{i}{\hbar}[H_s+H_{\mathrm{LS}},\varrho]
\\
&+\sum_{i,j=A,B}\frac{\gamma_{ij}}{2}
\left(2\sigma_j^-\varrho\sigma_i^+-\{\sigma_i^+\sigma_j^-,\varrho\}\right).
\end{aligned}
\label{eq:master}
\end{equation}
Here \(\varrho\) is the two-probe density operator, with \(\sigma_i^\pm\) as defined in Eq.~\eqref{eq:sigma_def}. We use \(\gamma_{ij}\) for the full decay matrix and \(\delta\gamma_{ij}\) for its scattering correction. The single-probe free-space Lamb shifts are absorbed into \(H_s\), while the off-diagonal free-space exchange is bias independent at fixed \(\omega_{\mathrm{pr}}\) and cancels in the ON/OFF difference; the remaining scattering part is
\begin{equation}
H_{\mathrm{LS}}^{\mathrm{sc}}
=
\hbar\sum_{i,j=A,B}
\Omega_{ij}^{\mathrm{sc}}\,
\sigma_i^+\sigma_j^- .
\label{eq:HLS}
\end{equation}
In the FMR-dominated bias window of the scalar benchmark, the susceptibility's narrowest spectral scale is the Gilbert-broadened linewidth, of order \(\alpha\omega_{\mathrm{pr}}\). Since \(\alpha\omega_{\mathrm{pr}}/2\pi\simeq0.18\,\mathrm{GHz}\) exceeds the mHz--Hz collective rates found below by many orders of magnitude, the Green-tensor spectrum is flat on the scale of these rates and the kernels may be evaluated at the fixed frequency \(\omega_{\mathrm{pr}}\). At low temperature, \(\gamma_{ij}=\gamma^{\downarrow}_{ij}\). Setting \(\nth+1\to1\), subtracting the free-space part of Eq.~\eqref{eq:gamma_down}, and projecting onto \(\mathbf m_i=m_c\hat{\mathbf z}\) give the scattering contribution for the \(zz\) benchmark,
\begin{equation}
\delta\gamma_{ij}^{(zz)}=
\frac{2\mu_0m_c^2}{\hbar}
\Im \mathcal G^{\mathrm{sc}}_{BB,zz}(\mathbf r_i,\mathbf r_j;\omega_{\mathrm{pr}},B).
\label{eq:deltagamma}
\end{equation}
For \(i=j\), the same expression gives the local scattering correction. Separating the free-space tensor from the coincident-point kernel defines the finite cavity-induced correction
\begin{equation}
\mathcal G^{\mathrm{sc}}_{BB}(\mathbf r,\mathbf r;\omega,B)=
\mathcal G^{\mathrm{cav}}_{BB}(\mathbf r,\mathbf r;\omega,B)-
\mathcal G^{(0)}_{BB}(\mathbf r,\mathbf r;\omega).
\label{eq:self_regularized}
\end{equation}
Here \(\mathcal G^{\mathrm{cav}}_{BB}\) is the full Green tensor of the two-film cavity. Subtracting the universal free-space contribution, including its singular real self-field, leaves a finite coincident-point scattering correction whose imaginary and real parts give the cavity-induced local linewidth and frequency shift, respectively.

For reciprocal channels, the scattering contribution to the coherent exchange coupling entering Eq.~\eqref{eq:HLS} is fixed by the real part of the projected retarded scattering tensor,
\begin{equation}
\Omega^{\mathrm{sc}}_{ij}=
-\frac{\mu_0}{\hbar}
\mathbf m_i^*\cdot
\Re \Gbb^{\mathrm{sc}}(\mathbf r_i,\mathbf r_j;\omega_{\mathrm{pr}},B)
\cdot\mathbf m_j.
\label{eq:OmegaReG_general}
\end{equation}
The standard macroscopic-QED resonant dipole--dipole Hamiltonian is written as \(-\hbar\delta_{ij}\sigma_i^+\sigma_j^-\), where the coupling parameter \(\delta_{ij}\) is proportional to the real retarded Green tensor~\cite{DungKnollWelsch2002}. Therefore, the \(+\hbar\Omega_{ij}^{\mathrm{sc}}\sigma_i^+\sigma_j^-\) convention in Eq.~\eqref{eq:HLS} requires the overall minus sign in Eq.~\eqref{eq:OmegaReG_general}.
For the \(zz\) benchmark, Eq.~\eqref{eq:OmegaReG_general} gives
\begin{equation}
\Omega^{\mathrm{sc},zz}_{ij}=
-\frac{\mu_0m_c^2}{\hbar}
\Re \mathcal G^{\mathrm{sc}}_{BB,zz}(\mathbf r_i,\mathbf r_j;\omega_{\mathrm{pr}},B).
\label{eq:OmegaReG}
\end{equation}
For \(\mathbf r_i\neq\mathbf r_j\), any projection \(\mathcal G\) of the retarded off-diagonal scattering tensor formed with frequency-independent coefficients is analytic in the upper half-plane by causality. For real projection coefficients, the reality of the underlying time-domain response also gives the Schwarz relation \(\mathcal G(-\omega^*)=\mathcal G^*(\omega)\). If the contribution from the contour at infinity vanishes, its real and imaginary parts obey the unsubtracted Kramers--Kronig relation
\begin{equation}
\Re \mathcal G(\omega_{\mathrm{pr}})=
\frac{2}{\pi}\mathcal P\int_0^\infty \dd\omega\,
\frac{\omega\,\Im\mathcal G(\omega)}{\omega^2-\omega_{\mathrm{pr}}^2}.
\label{eq:KK}
\end{equation}
If that asymptotic condition fails, an appropriately subtracted dispersion relation is required. Thus \(\Omega_{12}^{\mathrm{sc}}\) and \(\delta\gamma_{12}\) are determined by the dispersive and dissipative parts, respectively, of the same causal scattering response.

For identical probes at symmetry-equivalent positions in the reciprocal scalar channel, the decay matrix has equal diagonal elements, \(\gamma_{AA}=\gamma_{BB}\equiv\gamma_{11}\), and equal real off-diagonal elements, \(\gamma_{AB}=\gamma_{BA}\equiv\gamma_{12}\). It is therefore diagonal in the symmetric and antisymmetric basis, with
\begin{equation}
\Gamma_\pm=\gamma_{11}\pm\gamma_{12},
\qquad
\Gamma_+-\Gamma_-=2\gamma_{12}.
\label{eq:linewidth_split}
\end{equation}
The sign of \(\delta\gamma_{12}\) determines which collective mode has the smaller cavity-induced linewidth correction: the symmetric mode for \(\delta\gamma_{12}<0\) and the antisymmetric mode for \(\delta\gamma_{12}>0\). Although the scattering correction is signed, the full decay matrix constructed from \(\Gbb^{(0)}+\Gbb^{\mathrm{sc}}\) remains positive semidefinite for a passive reservoir~\cite{ScheelBuhmann2008,Buhmann2012}.

\subsection{Collective bright modes and finite-volume averaging}
\label{subsec:bright}

The single-pair kernels above define the microscopic common-bath matrix elements. In an ensemble implementation, the experimentally accessible excitation is a collective bright mode~\cite{Dicke1954,GrossHaroche1982,Ruostekoski2023Cooperative}.

For the uniform mode, the single-excitation bright state of each ensemble is the symmetric Dicke state
\begin{equation}
|W_A\rangle=\frac{1}{\sqrt{N_A}}\sum_{a\in A}|e_a\rangle,
\qquad
|W_B\rangle=\frac{1}{\sqrt{N_B}}\sum_{b\in B}|e_b\rangle.
\label{eq:wstates}
\end{equation}
Here, \(|e_a\rangle\) denotes the product state with spin \(a\) excited and all other spins in ensemble \(A\) in their ground states, and similarly for \(|e_b\rangle\). The corresponding bright-mode kernel is the coherent sum of the pairwise kernels,
\begin{equation}
\mathcal K^{\mathrm{br}}_{AB}=
\sum_{a\in A}\sum_{b\in B}
c_{A,a}^*c_{B,b}\,\mathcal K_{ab},
\label{eq:bright_discrete}
\end{equation}
where \(c_{\nu,i}\) are the bright-mode amplitudes, equal to \((N_\nu)^{-1/2}\) for the uniform states of Eq.~\eqref{eq:wstates}. When the participating spins are described by smooth mode envelopes, this double sum becomes a double volume integral. Let \(n_\nu(\mathbf r)\) be the spin number density of ensemble \(\nu=A,B\) and \(c_\nu(\mathbf r)\) the corresponding single-excitation bright-mode amplitude, normalized by
\begin{equation}
\int_{V_\nu}\dd^3r\,n_\nu(\mathbf r)|c_\nu(\mathbf r)|^2=1 .
\label{eq:bright_norm}
\end{equation}
In this continuum limit,
\begin{equation}
\begin{aligned}
\mathcal K^{\mathrm{br}}_{AB}
={}&\int_{V_A}\dd^3r\int_{V_B}\dd^3r'\,
n_A(\mathbf r)n_B(\mathbf r')\,
c_A^*(\mathbf r)c_B(\mathbf r')
\\
&\times
\mathbf m_A^*\cdot\Gbb(\mathbf r,\mathbf r';\omega_{\mathrm{pr}},B)\cdot\mathbf m_B .
\end{aligned}
\label{eq:bright_average}
\end{equation}
Let \(N_\nu^{\mathrm{eff}}\) denote the effective number of spins participating in mode \(\nu\), after orientation selection, inhomogeneous broadening, spectral filtering, and internal disorder. In the ideal bright-mode limit, all participating pairs sample the same single-pair kernel and share a common spatial phase and transition polarization. With uniform amplitudes \(c_\nu=1/\sqrt{N_\nu^{\mathrm{eff}}}\), Eq.~\eqref{eq:bright_average} then gives
\begin{equation}
\gamma_{AB}^{\mathrm{br}}\simeq\sqrt{N_A^{\mathrm{eff}}N_B^{\mathrm{eff}}}\,\gamma_{12},
\qquad
\Omega_{AB}^{\mathrm{br}}\simeq\sqrt{N_A^{\mathrm{eff}}N_B^{\mathrm{eff}}}\,\Omega_{12}.
\label{eq:bright_scaling}
\end{equation}
For ensemble \(\nu=A,B\), the bright-mode decay rate is the double sum
\begin{equation}
\Gamma_\nu^{\mathrm{br}}=\sum_{i,j\in\nu}c_{\nu,i}^*c_{\nu,j}\gamma_{ij}.
\label{eq:bright_decay_sum}
\end{equation}
The sum contains \(N_\nu^{\mathrm{eff}}\) self terms, whose normalized contribution is the single-spin rate \(\gamma_{11}\), and \(N_\nu^{\mathrm{eff}}(N_\nu^{\mathrm{eff}}-1)\) off-diagonal intra-ensemble terms. With the same uniform amplitudes,
\begin{equation}
\Gamma_\nu^{\mathrm{br}}\simeq\gamma_{11}+\bigl(N_\nu^{\mathrm{eff}}-1\bigr)\bar\gamma_{\mathrm{intra},\nu},
\label{eq:bright_diagonal}
\end{equation}
where \(\bar\gamma_{\mathrm{intra},\nu}\) is the mode-averaged intra-ensemble decay kernel.
For an NV ensemble with equal populations in the four crystallographic orientations, selecting one orientation reduces the participating population to one quarter before spectral and spatial filtering~\cite{Barry2020}.

\section{Numerical implementation and benchmark results}
\label{sec:numerics}
\begin{figure}[t]
\centering
\includegraphics[width=\columnwidth]{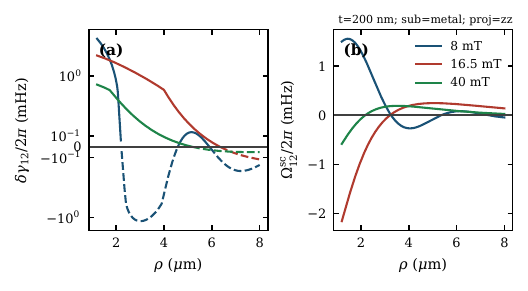}
\caption{Single-pair finite-frequency kernels for the \(zz\) projection of the finite-thickness scalar cavity integral. The \(zz\) projection is the component benchmark used to match the static TE normalization of Ref.~\cite{Kwon2026Magnetostatic}. Panel (a) shows the signed off-diagonal scattering kernel \(\delta\gamma_{12}^{(zz)}/2\pi\) on a symmetric logarithmic scale; solid and dashed segments denote positive and negative values. Panel (b) shows the scattering contribution to the coherent exchange coupling, \(\Omega_{12}^{\mathrm{sc},zz}/2\pi\). For the cavity-induced scattering contribution, the sign in panel (a) determines which collective mode has the smaller linewidth correction: the symmetric mode for \(\delta\gamma_{12}^{(zz)}<0\) and the antisymmetric mode for \(\delta\gamma_{12}^{(zz)}>0\).}
\label{fig:singlekernels}
\end{figure}

Having defined the common-bath kernels for both a single probe pair and ensemble bright modes, we evaluate their magnitude and sign as functions of bias and probe separation using the finite-thickness slab response. We compute Eq.~\eqref{eq:planarGBB_scalar} numerically in terms of the dimensionless variable \(q=\kb L\) and truncate the integral at \(q_{\max}=80\). The image terms are suppressed by \(e^{-q}\) or faster, making the omitted tail negligible. In the constant-reflection limit, the \(\omega\to0\) integral reproduces the real-space image series of Appendix~\ref{app:staticlimit} to better than \(10^{-6}\) over the plotted separations. Details of the integration grids, retarded-branch implementation, and thermal-factor approximation are given in Appendix~\ref{app:numerics}.

The numerical benchmarks use a representative Ni-like parameter set based on room-temperature material data, with \(T=20\,\mathrm{mK}\) fixing the probe thermal occupation. The conductivity controls the slab response and local loss, \(\alpha\) sets the FMR linewidth, \(M_s\) sets the susceptibility amplitude, and \(M_{\mathrm{eff}}\) fixes the resonance field. The value \(\mu_0M_s=0.617\,\mathrm T\) is taken from Ref.~\cite{Crangle1971}. These material parameters are sample dependent, and ferromagnetic relaxation can also vary with temperature~\cite{BhagatLubitz1974}. Values measured for a given sample at its operating temperature can be substituted without changing the Green-tensor construction.

Figure~\ref{fig:singlekernels} shows the single-pair scattering kernels \(\delta\gamma_{12}^{(zz)}\) and \(\Omega_{12}^{\mathrm{sc},zz}\) for the main \(zz\) benchmark. The component-normalized transverse-average projection, in which the odd- and even-reflection branches enter with opposite relative signs, is evaluated in Appendix~\ref{app:transverseprojection}.

\begin{figure}[t]
\centering
\includegraphics[width=\columnwidth]{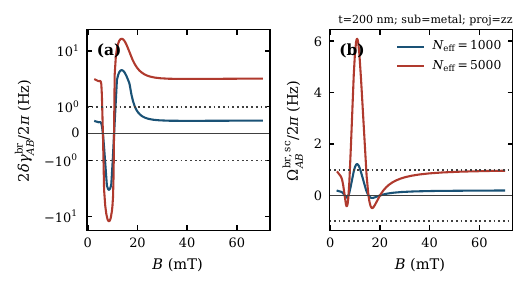}
\caption{Ideal uniform bright-mode estimate retaining only the cavity-induced scattering contributions at \(\rho=3\,\mu\mathrm m\), shown for two values of the effective number of participating spins. Panel (a) shows the cavity-scattering contribution \(2\delta\gamma_{AB}^{\mathrm{br},zz}/2\pi\) to the bright-mode linewidth splitting, with dotted horizontal lines marking the \(\pm1\,\mathrm{Hz}\) reference level. Panel (b) shows the cavity-induced bright-mode coherent exchange coupling \(\Omega_{AB}^{\mathrm{br,sc},zz}/2\pi\) for the same effective numbers of participating spins.}
\label{fig:bright}
\end{figure}

Applying Eq.~\eqref{eq:bright_scaling} to the scattering part of each single-pair kernel gives the ideal collective scales shown in Fig.~\ref{fig:bright}. For the \(zz\) projection with \(N_A^{\mathrm{eff}}=N_B^{\mathrm{eff}}=N_{\mathrm{eff}}\), the cavity-induced bright-mode linewidth splitting and coherent exchange coupling are \(2\delta\gamma_{AB}^{\mathrm{br},zz}=2N_{\mathrm{eff}}\delta\gamma_{12}^{(zz)}\) and \(\Omega_{AB}^{\mathrm{br,sc},zz}=N_{\mathrm{eff}}\Omega_{12}^{\mathrm{sc},zz}\). For the full decay matrix, with \(\bar\gamma_{\mathrm{intra},A}=\bar\gamma_{\mathrm{intra},B}=\bar\gamma_{\mathrm{intra}}\), Eqs.~\eqref{eq:bright_scaling} and \eqref{eq:bright_diagonal} give a splitting-to-linewidth ratio that approaches the \(N_{\mathrm{eff}}\)-independent value \(2|\gamma_{12}(\rho)|/\bar\gamma_{\mathrm{intra}}\) at large \(N_{\mathrm{eff}}\).

For pointlike probes, we compare the off-diagonal and local scattering corrections using \(\delta\gamma_{11}^{(zz)}\) as the local reference scale. At the resonant bias and \(\rho=3\,\mu\mathrm m\), \(2|\delta\gamma_{12}^{(zz)}|/\delta\gamma_{11}^{(zz)}\simeq0.66\), showing that the off-diagonal contribution to the linewidth splitting is comparable to the local scattering correction. For a finite ensemble, the corresponding inter- and intra-ensemble kernels must be averaged over the relevant mode volumes using Eqs.~\eqref{eq:bright_average} and \eqref{eq:bright_diagonal}.

The plotted curves correspond to \(N_{\mathrm{eff}}=10^3\) and \(5\times10^3\), for which the cavity-induced linewidth splitting and coherent exchange coupling grow linearly with \(N_{\mathrm{eff}}\) and reach the hertz range. The experimentally relevant figure of merit is the ratio of this differential splitting to the measured bright-mode linewidth. The decay contribution to that linewidth is given by Eq.~\eqref{eq:bright_diagonal}; the measured value may also include sample-dependent dephasing.

\begin{table*}[t]
\caption{Parameters used in the finite-thickness scalar TE calculation.}
\label{tab:params}
\centering
\renewcommand{\arraystretch}{1.13}
\setlength{\tabcolsep}{6pt}
\resizebox{\textwidth}{!}{%
\begin{tabular}{lll}
\toprule
\textbf{Quantity} & \textbf{Value} & \textbf{Role} \\
\midrule
Probe frequency \(\omega_{\mathrm{pr}}/2\pi\) & \(2.87\,\mathrm{GHz}\) & NV-like transition \\
Temperature \(T\) & \(20\,\mathrm{mK}\) & low thermal occupation \\
Gap width \(L\) & \(3\,\mu\mathrm m\) & cavity scale \\
Probe heights & \(z_A=z_B=L/2\) & symmetric placement \\
Film thickness \(t\) & \(200\,\mathrm{nm}\) & finite slab benchmark \\
Substrate model & nonmagnetic conductor, \(\sigma=\sigma_{\mathrm{Ni}}\) & conducting backing layer \\
Projection axis & main: \(zz\); appendix: \((xx+yy)/2\) & selected transition-moment projection \\
Ni-like \(\mu_0M_s\) & \(0.617\,\mathrm T\) & bulk-Ni value at \(293\,\mathrm K\), sets \(\omega_M\)~\cite{Crangle1971} \\
Effective magnetization \(\mu_0M_{\mathrm{eff}}\) & \(0.617\,\mathrm T\) &
illustrative identification with \(M_s\); sets the resonance field \\
Gyromagnetic ratio \(\gamma_{\mathrm F}/2\pi\) & \(28.0\,\mathrm{GHz/T}\) &
FMR conversion factor \\
Gilbert damping \(\alpha\) & \(0.064\) & room-temperature Ni FMR damping~\cite{Oogane2006Damping} \\
Conductivity \(\sigma\) & \(1.43\times10^7\,\mathrm{S/m}\) & room-temperature bulk value, illustrative input \\
Component transition moment \(m_c\) & \(g_e\mu_B/\sqrt{2}\) & Cartesian spin-flip matrix element \\
\bottomrule
\end{tabular}
}
\end{table*}

\section{Signal discrimination and benchmark scope}
\label{sec:discussion}

The numerical kernels determine the cavity-induced decay and exchange scales. Experimentally, these quantities appear as frequency and linewidth splittings of the symmetric and antisymmetric collective modes. We therefore express the calculated kernels as ON/OFF spectroscopic observables and show how to separate them from common-mode shifts and background contributions. Throughout this section, \(\gamma_{12}\) and \(\Omega_{12}\) denote the off-diagonal decay rate and exchange coupling for either two symmetry-equivalent probes or the two symmetry-equivalent ensemble bright modes of Eq.~\eqref{eq:wstates}. These bright-mode values are given by Eq.~\eqref{eq:bright_scaling}. For the probe pair, the common diagonal element is \(\gamma_{11}\); for the bright-mode pair, it is \(\Gamma_A^{\mathrm{br}}=\Gamma_B^{\mathrm{br}}\) as defined by Eq.~\eqref{eq:bright_diagonal}. Equation~\eqref{eq:linewidth_split} applies in either case after replacing \(\gamma_{11}\) by the corresponding common diagonal element. The relation \(\Gamma_+-\Gamma_-=2\gamma_{12}\) gives the differential common-bath linewidth response
\begin{equation}
\Delta(\Gamma_+-\Gamma_-)=
2[\gamma_{12}(B_{\mathrm{on}})-\gamma_{12}(B_{\mathrm{off}})].
\label{eq:diff_linewidth}
\end{equation}
The off-diagonal coherent exchange coupling shifts the symmetric and antisymmetric modes by \(+\Omega_{12}\) and \(-\Omega_{12}\), respectively, giving a frequency separation \(\omega_+-\omega_-=2\Omega_{12}\). This separation is the spectroscopic signature of the collective Lamb shift. The ON/OFF change of the splitting is therefore
\begin{equation}
\Delta(\omega_+-\omega_-)=
2\big[\Omega_{12}(B_{\mathrm{on}})-\Omega_{12}(B_{\mathrm{off}})\big].
\label{eq:diff_shift}
\end{equation}
ON/OFF subtraction of the measured linewidth and frequency splittings removes bias-independent spectral offsets.

Although the two-mode comparison in Eq.~\eqref{eq:diff_shift} isolates the exchange contribution, the ON/OFF shift of either branch alone at fixed probe frequency is
\begin{equation}
\Delta\omega_\pm=
\Delta\Omega_{\mathrm{com}}\pm\Delta\Omega_{12},
\label{eq:single_mode_shift}
\end{equation}
where \(\Delta\Omega_{12}=\Omega_{12}(B_{\mathrm{on}})-\Omega_{12}(B_{\mathrm{off}})\) and \(\Delta\Omega_{\mathrm{com}}\) is the common-mode shift, equal to \(\Delta\Omega_{11}\) for the symmetry-equivalent probe pair and to the corresponding diagonal collective shift for the bright-mode pair. A single resolved branch therefore does not determine \(\Delta\Omega_{12}\) unless this common-mode contribution is removed independently. The exchange term follows instead from one half of the ON/OFF change in the mode splitting, or equivalently by subtracting the measured midpoint shift from either branch.

\subsection{Proposed measurement protocol}
\label{subsec:protocol}

A practical protocol measures the collective spectra at \(B_{\mathrm{on}}\) and \(B_{\mathrm{off}}\) and fits the two mode frequencies \(\omega_\pm\) and linewidths \(\Gamma_\pm\) at each bias setting. The ON/OFF changes in the frequency and linewidth splittings then give the signals in Eqs.~\eqref{eq:diff_shift} and \eqref{eq:diff_linewidth}. As an additional diagnostic, the measurement may be repeated under \(B\to-B\) in geometries that permit field reversal, thereby identifying bias-odd background contributions. The probe transition frequency is recorded during the bias sweep, and the local linewidths are measured at the same bias settings.

The same magnetic field used to tune the ferromagnet can also shift an NV-center probe transition. In the regime where transverse-state mixing is negligible, a selected transition is
\begin{equation}
\omega_\eta(B)=2\pi D+
\eta\gamma_e B_\parallel+
\delta_{\mathrm{strain}}+
\delta_{\mathrm{hf}}+\cdots,
\qquad \eta=\pm1,
\label{eq:nv_frequency}
\end{equation}
where \(D\simeq2.87\,\mathrm{GHz}\), \(\gamma_e=2\pi\times28\,\mathrm{GHz/T}\), \(B_\parallel\) is the projection of the applied field on the selected NV axis, and \(\delta_{\mathrm{strain}}\) and \(\delta_{\mathrm{hf}}\) denote strain and hyperfine corrections~\cite{Doherty2013}. An uncompensated bias scan therefore samples \(\Gbb[\omega_\eta(B),B]\), combining the material response with the bias-dependent probe frequency. The fixed-\(\omega_{\mathrm{pr}}\) benchmark instead requires the selected transition to remain at \(\omega_{\mathrm{pr}}\). Compensating the longitudinal Zeeman shift requires an independently controlled field component at the NV site that cancels \(B_\parallel\) without changing the bias applied to the ferromagnetic films. A transverse-bias geometry alone, however, does not keep the transition fixed, because the transverse Zeeman term mixes the \(m_s\) states and changes both \(\omega_\eta(B)\) and the transition moment \(\mathbf m_\eta(B)\). For transverse bias, these quantities follow from diagonalizing the full NV ground-state Hamiltonian before forming the projected Green-tensor kernel. The zero-field splitting \(D\) is also temperature dependent~\cite{Doherty2013,Acosta2010}; at the fixed operating temperature used here, that shift is absorbed into \(\omega_{\mathrm{pr}}\).

At \(T=20\,\mathrm{mK}\) the thermal occupation is \(\nth\simeq10^{-3}\). In NV ensembles, the two-phonon Raman and Orbach-type processes are strongly suppressed on cooling, and sample-dependent cross-relaxation dominates the low-temperature longitudinal relaxation~\cite{Jarmola2012}. The measured diagonal linewidth contains the cavity correction together with sample-dependent relaxation, dephasing, and inhomogeneous broadening. Comparison of the measured diagonal linewidth with the calculated \(\delta\gamma_{11}^{(zz)}\) therefore constrains the noncavity contribution. The visibility of a resolved bright-mode splitting additionally depends on the mode-averaged intra-ensemble decay kernel in Eq.~\eqref{eq:bright_diagonal}.

Estimating the ON/OFF change of the half-splitting is a frequency-estimation problem analogous to the static TE protocol~\cite{Kwon2026Magnetostatic}. For a Ramsey measurement with negligible dead time, readout efficiency \(C\), interrogation time \(T_R\), and averaging time \(\tau_{\mathrm{av}}\), the single-mode frequency uncertainty is
\[
\sigma_f\simeq \frac{1}{2\pi C\sqrt{T_R\tau_{\mathrm{av}}}},
\]
up to protocol-dependent numerical factors of order unity~\cite{Degen2017}. With four independently fitted mode frequencies of equal uncertainty \(\sigma_f\), the ON/OFF change of the half-splitting also carries uncertainty \(\sigma_f\), so resolving it requires \(\sigma_f\lesssim|\Delta\Omega_{12}|/(2\pi)\).

\subsection{Scope of the scalar calculation}
\label{subsec:scope}

The scalar local permeability of Eq.~\eqref{eq:polder} excludes spin-wave dispersion, while the polarization-diagonal TE treatment omits polarization conversion. Likewise, the scalar damping parameter \(\alpha\) neglects the wave-vector dependence of Gilbert damping that modifies magnon lifetimes in Ni and other elemental ferromagnets~\cite{Lu2023NonlocalDamping}.

For the mid-gap geometry, \(z_A=z_B=L/2\), the numerator of Eq.~\eqref{eq:planarGBB_scalar} reduces to \(\mathcal N=2r_s e^{-\kappa_0L}+2r_s^2e^{-2\kappa_0L}\). Since \(\kappa_0\simeq k_\perp\) in the near-field sector, these exponential factors suppress contributions with \(k_\perp L\gg1\) and select a wave-vector window with \(k_\perp L=O(1)\). Whether spatial-dispersion corrections are negligible over this window is material dependent. For a laterally homogeneous stack, the planar angular-spectrum representation remains applicable, but the reflection amplitude must then be obtained from the coupled electromagnetic--magnetization boundary problem, in which exchange stiffness and surface spin pinning enter together~\cite{StancilPrabhakar2009}. Polarization conversion, where present, additionally requires the full reflection matrix. Inserting \(\mu_\perp(\omega,k_\perp,B)\) into the local Fresnel and slab formulas of Eqs.~\eqref{eq:rab_te} and \eqref{eq:rslab} does not in general reproduce that amplitude.

\section{Conclusion}
\label{sec:conclusion}

We have formulated the finite-frequency magnetic common-bath response of a ferromagnetic planar cavity in terms of real-frequency magnetic Green tensors, with a normalization directly comparable to the static TE coupling-frequency shift. The finite-thickness reciprocal TE benchmark, built from a Polder-type scalar permeability, gives bias-tunable off-diagonal decay rates and coherent exchange couplings for two mid-gap spin probes. In the constant-reflection limit, the \(\omega\to0\) Green-tensor kernel reduces to the magnetostatic response; for the finite slab, the same limit retains the static TE reflection amplitude and fixes the normalization shared with the companion static calculation. Within the scattering-sector benchmark, \(2|\delta\gamma_{12}^{(zz)}|/\delta\gamma_{11}^{(zz)}\simeq0.66\) at the resonant bias and \(\rho=3\,\mu\mathrm m\), while ideal uniform bright-mode estimates with \(N_{\mathrm{eff}}=10^3\) and \(5\times10^3\) yield corresponding cavity-induced splitting and shift scales in the hertz range.

In the millikelvin GHz regime, the linewidth splitting, collective Lamb shift, and local linewidths probe the body-assisted magnetic reservoir, whose fluctuations are shaped by the ferromagnetic films and the cavity geometry. Together, \(\Delta f\), \(\gamma_{12}\), and \(\Omega_{12}\) characterize the static, dissipative, and dispersive components of the magnetic boundary response within a single Green-tensor normalization. The same formalism extends to biased gyrotropic films by replacing the scalar TE amplitude with the full TE/TM reflection matrix and forming the projected kernels from the generalized real and imaginary parts appropriate to nonreciprocal media. The scalar reciprocal calculation thereby provides an ON/OFF extraction protocol for magnetic common-bath measurements in ferromagnetic planar cavities.


\section*{Data Availability}
The data that support the findings of this article are available from the author upon reasonable request.

\appendix
\section{Free-space normalization}
\label{app:normalization}

The prefactors in the open-system rate expressions are fixed by the free-space magnetic-dipole decay rate \(\Gamma_m^{(0)}=\mu_0m^2k_0^3/(3\pi\hbar)\). Rotational invariance makes the coincident-point value of \(\Im\Gbb^{(0)}\) isotropic. Substitution into \(\Gamma_m=(2\mu_0/\hbar)\,\mathbf m\cdot\Im\Gbb^{(0)}\cdot\mathbf m\) then gives Eq.~\eqref{eq:free_norm_intro}. This convention separates the universal free-space self term, including the singular real self-field, from the finite cavity-induced scattering correction defined in Eq.~\eqref{eq:self_regularized}.

\section{Static-limit image series}
\label{app:staticlimit}

For a \(k\)-independent static TE reflection amplitude \(r=(\mu-1)/(\mu+1)\), the limit \(\omega\to0\) gives \(\kappa_0\to k_\perp\). Writing \(k\equiv k_\perp\), Eq.~\eqref{eq:planarGBB_scalar} reduces to
\begin{equation}
\begin{aligned}
&\mathcal G^{\mathrm{sc}}_{BB,zz}(\rho,z_A,z_B;0,B)
\\
&\qquad=\frac{1}{4\pi}\int_0^\infty\dd k\,k^2J_0(k\rho)
\frac{\mathcal N_0(k)}{1-r^2e^{-2kL}},
\end{aligned}
\label{eq:static_integral}
\end{equation}
with
\begin{equation}
\begin{aligned}
\mathcal N_0(k)=&
r\left[e^{-k\Sigma_z}+e^{-k(2L-\Sigma_z)}\right]
\\
&+r^2\left[e^{-k(2L-\Delta z)}+e^{-k(2L+\Delta z)}\right].
\end{aligned}
\end{equation}
Expanding the cavity denominator as
\begin{equation}
\frac{1}{1-r^2e^{-2kL}}=\sum_{n=0}^{\infty}r^{2n}e^{-2nkL}
\label{eq:denominator_expansion}
\end{equation}
converts each power of \(r^2e^{-2kL}\) into one additional round trip between the two interfaces and shifts each image height by \(2nL\). The standard Hankel-transform identity
\begin{equation}
\begin{aligned}
\mathcal F(a,\rho)
&=\int_0^\infty\dd k\,k^2e^{-ka}J_0(k\rho)
\\
&=\frac{2a^2-\rho^2}{(a^2+\rho^2)^{5/2}},
\qquad a>0,
\end{aligned}
\label{eq:Fkernel}
\end{equation}
follows by differentiating \(\int_0^\infty\dd k\,e^{-ka}J_0(k\rho)=(a^2+\rho^2)^{-1/2}\)
twice with respect to \(a\). Applying it term by term after the denominator expansion yields the image series
\begin{equation}
\begin{aligned}
\mathcal G_{BB,zz}^{\mathrm{sc}}
&(\rho,z_A,z_B;0,B)
=\frac{1}{4\pi}\Bigg[
\sum_{n=0}^{\infty} r^{2n+1}
\Big\{
\mathcal F(\Sigma_z+2nL,\rho)
\\
&\qquad
+\mathcal F(2L-\Sigma_z+2nL,\rho)
\Big\}
\\
&\qquad
+\sum_{n=1}^{\infty} r^{2n}
\Big\{
\mathcal F(2nL-\Delta z,\rho)
\\
&\qquad
+\mathcal F(2nL+\Delta z,\rho)
\Big\}
\Bigg].
\end{aligned}
\label{eq:static_image_series_mathcal}
\end{equation}

\begin{figure}[t]
\centering
\includegraphics[width=\columnwidth]{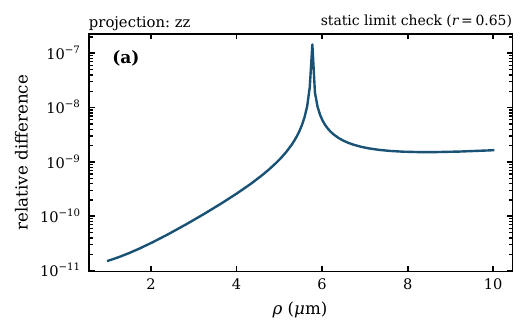}
\caption{Numerical static-limit check for a constant reflection coefficient \(r=0.65\). The \(k\)-space integral and the real-space image series agree to better than \(10^{-6}\) over the plotted separations, with the residual difference set by numerical quadrature.}
\label{fig:staticcheck}
\end{figure}

The powers \(r^{2n+1}\) and \(r^{2n}\) count odd- and even-reflection paths, respectively. Multiplication by \(\mu_0\) gives the SI magnetostatic Green tensor used for static frequency shifts. The finite-thickness static slab response is obtained by retaining \(\rslab(0,k_\perp,B;t)\) in the \(k\)-space integral. Figure~\ref{fig:staticcheck} compares the constant-\(r\) static \(k\)-integral and the image series numerically.

For \(\rho\gg L\) and fixed \(|r|<1\), the factors \(r^n\) make the image sum uniformly convergent. The large-distance limit can then be taken term by term, giving \(\mathcal F(a,\rho)\to-1/\rho^3\) for each fixed image height \(a\). The odd- and even-reflection branches contribute \(\sum_{n\ge0}r^{2n+1}=r/(1-r^2)\) and \(\sum_{n\ge1}r^{2n}=r^2/(1-r^2)\), whose sum is \(r/(1-r)\). Each brace in Eq.~\eqref{eq:static_image_series_mathcal} holds two image terms, which doubles the prefactor \(1/4\pi\). The resulting algebraic tail is
\begin{equation}
\mu_0\mathcal G_{BB,zz}^{\mathrm{sc}}(\rho;0,B)=
-\frac{\mu_0 r}{2\pi(1-r)}\frac{1}{\rho^3}
\left[1+O\left(\frac{L^2}{\rho^2}\right)\right],
\label{eq:large_rho_tail}
\end{equation}
which matches the constant-\(r\) image-series reference of the companion magnetostatic calculation~\cite{Kwon2026Magnetostatic}.

\section{Transverse-average projection and reflection parity}
\label{app:transverseprojection}

The main text uses the \(zz\) projection as the scalar benchmark to match the static TE normalization. With \(\hat{\boldsymbol\mu}=-g_e\mu_B\hat{\mathbf S}/\hbar\) and \(|\langle0|\hat S_x|\pm1\rangle|=|\langle0|\hat S_y|\pm1\rangle|=\hbar/\sqrt2\), the \(m_s=0\leftrightarrow\pm1\) transition has Cartesian matrix elements
\begin{equation}
|\langle0|\hat\mu_x|\pm1\rangle|
=|\langle0|\hat\mu_y|\pm1\rangle|
=\frac{g_e\mu_B}{\sqrt2}
\equiv m_c .
\label{eq:nv_component_moment}
\end{equation}
For an NV axis normal to the films, the transition moments may be written, up to an overall phase, as \(\mathbf m_\pm=m_c(\hat{\mathbf x}\pm i\hat{\mathbf y})\).
We choose the \(x\) axis along \(\bm\rho_A-\bm\rho_B\). Reflection symmetry under \(y\to-y\) in the reciprocal scalar channel gives \(\mathcal G^{\mathrm{sc}}_{BB,xy}=\mathcal G^{\mathrm{sc}}_{BB,yx}=0\), and hence
\begin{equation}
\mathbf m_\pm^*\cdot\Gbb^{\mathrm{sc}}\cdot\mathbf m_\pm
=m_c^2\left(\mathcal G^{\mathrm{sc}}_{BB,xx}+\mathcal G^{\mathrm{sc}}_{BB,yy}\right).
\label{eq:nv_circular_projection}
\end{equation}
Dividing the circular contraction in Eq.~\eqref{eq:nv_circular_projection} by two gives the transverse average normalized per Cartesian component, \(m_c^2\mathcal G^{\mathrm{sc}}_{BB,\perp}\), where \(\mathcal G^{\mathrm{sc}}_{BB,\perp}=(\mathcal G^{\mathrm{sc}}_{BB,xx}+\mathcal G^{\mathrm{sc}}_{BB,yy})/2\). In the reciprocal scalar TE channel, its TE contribution is
\begin{equation}
\begin{aligned}
\mathcal G^{\mathrm{sc}}_{BB,\perp}
={}&\frac{1}{2\pi}\int_0^\infty\dd \kb\,\kb\,
\frac{\kappa_0}{4}J_0(\kb\rho)
\frac{\mathcal N_\perp}
{1-r_s^2e^{-2\kappa_0L}},\\
\mathcal N_\perp
={}&r_s\left[e^{-\kappa_0\Sigma_z}
+e^{-\kappa_0(2L-\Sigma_z)}\right]\\
&-r_s^2\left[e^{-\kappa_0(2L-\Delta z)}
+e^{-\kappa_0(2L+\Delta z)}\right].
\end{aligned}
\label{eq:planarGBB_transverse}
\end{equation}

The in-plane component of the TE magnetic polarization changes sign between upward and downward branches, whereas its normal component does not. Odd- and even-reflection paths therefore enter the transverse projection with opposite relative signs. Additional complete round trips preserve reflection parity, so the cavity denominator is the same as in Eq.~\eqref{eq:planarGBB_scalar}. Within a polarization-diagonal reciprocal stack, the full transverse projection also receives a contribution from the diagonal TM channel. In the dominant evanescent sector, the TM polarization prefactor is suppressed relative to the TE prefactor by \((k_0/\kappa_0)^2\). Equation~\eqref{eq:planarGBB_transverse} therefore gives the dominant TE contribution to the transverse projection. The corresponding transverse scattering kernels \(\delta\gamma_{ij}^{(\perp)}\) and \(\Omega_{ij}^{\mathrm{sc},\perp}\) follow from Eqs.~\eqref{eq:deltagamma} and \eqref{eq:OmegaReG}, respectively, by replacing \(\mathcal G^{\mathrm{sc}}_{BB,zz}\) with \(\mathcal G^{\mathrm{sc}}_{BB,\perp}\).

\begin{figure}[t]
\centering
\includegraphics[width=\columnwidth]{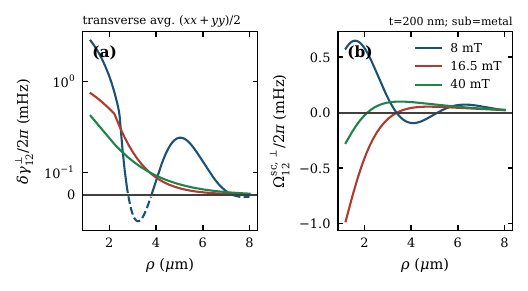}
\caption{Component-normalized transverse-average benchmark for the finite-thickness scalar cavity integral. Panel (a) shows the signed off-diagonal scattering kernel \(\delta\gamma_{12}^{(\perp)}/2\pi\) for the projection \(m_c^2(\mathcal G^{\mathrm{sc}}_{BB,xx}+\mathcal G^{\mathrm{sc}}_{BB,yy})/2\) on a symmetric logarithmic scale; solid and dashed segments denote positive and negative values, respectively. Panel (b) shows \(\Omega_{12}^{\mathrm{sc},\perp}/2\pi\) for the same projection. Odd- and even-reflection paths enter this projection with opposite relative signs, so the transverse curves are not fixed rescalings of the \(zz\) benchmark.}
\label{fig:transverseprojection}
\end{figure}

Figure~\ref{fig:transverseprojection} shows the transverse kernels obtained from Eq.~\eqref{eq:planarGBB_transverse} using the same finite-thickness slab amplitude and benchmark parameters as the main \(zz\) calculation. At the resonant bias, \(\delta\gamma_{12}^{(\perp)}\) remains positive throughout the plotted separation range, while the \(zz\) kernel changes sign. The comparison therefore isolates the reflection-parity dependence of the projected scalar TE response.

\section{Numerical benchmark details}
\label{app:numerics}

For the numerical evaluation of Eq.~\eqref{eq:planarGBB_scalar}, we use a uniform grid in \(q=\kb L\). All finite-frequency kernel figures use \(q_{\min}=10^{-4}\), \(q_{\max}=80\), and \(N_q=6000\) grid points. For the static constant-reflection check in Fig.~\ref{fig:staticcheck}, we use a separate grid with \(q_{\min}=10^{-6}\), \(q_{\max}=200\), and \(N_q=50000\). The retarded branch prescription of Eq.~\eqref{eq:kappa_branch} is used throughout. For \(\omega_{\mathrm{pr}}/2\pi=2.87\,\mathrm{GHz}\) and \(L=3\,\mu\mathrm m\), \(k_0L=1.8\times10^{-4}\), whereas the dominant near-field contribution is concentrated at \(q=O(1)\). The grid excludes \(0<q<q_{\min}\) and does not separately resolve \(q_{\min}<q<k_0L\). This propagating sector has phase-space weight of order \((k_0L)^3\simeq5.9\times10^{-12}\), leaving the displayed observables dominated by the evanescent near field.

For the plotted rates, we set \(\nth+1=1\). The value \(\nth+1=1.001\) from Eq.~\eqref{eq:nth_value} changes these rates by \(0.1\%\). The main \(zz\) results use \(t=200\,\mathrm{nm}\) and an idealized nonmagnetic conducting substrate with conductivity matched to the Ni-like film, \(\sigma=\sigma_{\mathrm{Ni}}\). Figure~\ref{fig:transverseprojection} uses the same material and geometric parameters with the component-normalized transverse average \((xx+yy)/2\).

\bibliographystyle{apsrev4-2}
\bibliography{refs}

\end{document}